\documentclass[letterpaper,twocolumn,fleqn]{article} 

\usepackage{ist}
\usepackage{graphicx}
\usepackage{caption}
\usepackage{subcaption}
\usepackage{amsmath} 
\usepackage{amssymb}

\usepackage{afterpage,lipsum,capt-of,graphicx}

\title{SLADS-Net: Supervised Learning Approach for Dynamic Sampling using Deep Neural Networks}

\author{ Yan Zhang$^{1}$, G. M. Dilshan Godaliyadda$^{2}$, Nicola Ferrier$^{3}$, Emine B. Gulsoy$^{4}$, Charles A. Bouman$^{2}$, Charudatta Phatak$^{1,*}$ \\
$^{1}$Materials Science Division, Argonne National Laboratory, Lemont, IL\\
$^{2}$ECE Department, Purdue University, West Lafayette, IN\\
$^{3}$Mathematics and Computer Science Division, Argonne National Laboratory, Lemont, IL\\
$^{4}$Department of Materials Science and Engineering, Northwestern University, Evanston, IL
}

\date{} 

\begin{document} 

\maketitle 

\thispagestyle{empty} 


\begin{abstract}
In scanning microscopy based imaging techniques, there is a need to develop novel data acquisition schemes that can reduce the time for data acquisition and minimize sample exposure to the probing radiation. Sparse sampling schemes are ideally suited for such applications where the images can be reconstructed from a sparse set of measurements. In particular, dynamic sparse sampling based on supervised learning has shown promising results for practical applications. However, a particular drawback of such methods is that it requires training image sets with similar information content which may not always be available. 

In this paper, we introduce a Supervised Learning Approach for Dynamic Sampling (SLADS) algorithm that uses a deep neural network based training approach. We call this algorithm SLADS-Net. We have performed simulated experiments for dynamic sampling using SLADS-Net in which the training images either have similar information content or completely different information content, when compared to the testing images. We compare the performance across various methods for training such as least-squares, support vector regression and deep neural networks. From these results we observe that deep neural network based training results in superior performance when the training and testing images are not similar.
We also discuss the development of a pre-trained SLADS-Net that uses generic images for training. Here, the neural network parameters are pre-trained so that users can directly apply SLADS-Net for imaging experiments.
\end{abstract}

\section{Introduction}

In certain conventional point-wise imaging modalities, each pixel measurement can take up to a few seconds to acquire, which can translate to hours or even days for large images (e.g. 1024 $\times$ 1024 pixels). Furthermore, exposure to a highly focused electron or X-ray beam for extended periods of time may also damage the underlying sample. Therefore, minimizing the image acquisition time and radiation damage is of critical importance. 
Static sampling methods, such as random sampling, uniformly spaced sampling and low-discrepancy sampling methods have been widely studied and used \cite{bib1}\cite{bib2}\cite{bib7}. Recently, sampling techniques where previous measurements are used to adaptively select new sampling locations have been presented. These methods, known as dynamic sampling methods, have been shown to significantly outperform traditional static sampling methods \cite{bib3}\cite{bib4}\cite{bib5}\cite{bib8}\cite{bib9}. In the Supervised Learning Approach for Dynamic Sampling (SLADS) algorithm \cite{bib5}, which is one such algorithm, the goal is to select the measurements that maximizes the Expected Reduction in Distortion (ERD). To train SLADS, one needs corresponding pairs of features extracted from previous measurements and the reduction in distortion (RD) due to new measurements. Then, SLADS uses least-squares regression to learn the mapping from features to the ERD. The SLADS framework assumes that the training images and the test images (i.e. images of underlying test samples) are similar. Here, we address scenarios where training images and images of test objects are dissimilar by proposing an improvement on SLADS which we call SLADS-Net.

To apply SLADS on sparse sampling, it is assumed that users have historical data - previously collected images of similar underlying samples. However, this assumption is not always true, for example, users may have very limited knowledge of a sample which perhaps comes from other imaging modalities. Training SLADS using dissimilar images may result in even worse performance than random sampling. In this paper, we propose using nonlinear machine learning methods for training, such as deep neural networks, to solve this problem. The goal of SLADS-Net is to develop a training-free data acquisition method for users by demonstrating that SLADS-Net can work well even when the training and testing images are dissimilar.  

\section{Theoretical Method for SLADS}
Supervised learning approach for Dynamic Sampling (SLADS) was developed by Godaliyadda et al. \cite{bib4}\cite{bib5}.
The goal of greedy dynamic sampling, in general, is to find the measurement which, when added to the existing dataset, has the greatest effect on the expected reduction in distortion (ERD). In this paper, we assume that every measurement is scalar valued. It is important to note that SLADS has also been generalized for vector measurements by Zhang et al \cite{yz}\cite{yz2} for Energy-Dispersive Spectroscopy, and by Scarborough et al. \cite{bib6} for X-ray diffraction imaging.

First, we define the image of the underlying object we wish to measure as $X \in \mathbb{R}^N$, and the value of location $s$ as $X_s$. Now assume we have already measured $k$ pixels from this image. Then we can construct a measurement vector,
$$
Y^{(k)} = 
\left[ 
\begin{array}{c}
s^{(1)}, X_{s^{(1)}} \\
\vdots \\
s^{(k)}, X_{s^{(k)}} 
\end{array}
\right] \ .
$$
Using $Y^{(k)}$ we can then reconstruct an image $\hat{X}^{(k)}$.

Then, we can define the distortion between the ground-truth $X$ and the reconstruction $\hat{X}^{(k)}$ as $D \left(X,  \hat{X}^{(k)}\right)$. Here $D \left(X,  \hat{X}^{(k)}\right)$ can be any metric that accurately quantifies the difference between $X$ and $\hat{X}^{(k)}$. In this paper, we define $D \left(X,  \hat{X}^{(k)}\right)$ as,
\begin{equation}
D \left(X,  \hat{X}^{(k)}\right) = \displaystyle\sum\limits_{i=1}^N |X_i - \hat{X}^{(k)}_i|,
\end{equation}
since we are focused on SLADS for continuous images.

Assume we measure pixel location $s$, where $s \in \left\lbrace \Omega \setminus \mathcal{S} \right\rbrace$, where $\Omega$ is the set containing indices of all pixels, and $\mathcal{S}$ is the set containing pixel locations of all measured pixels. Then, if we define the reconstruction performed using previous measurements and a new measurement at location $s$ as $\hat{X}^{(k;s)}$, we can define the reduction in distortion (RD) that results from measuring $s$ as,
\begin{eqnarray}
R^{(k;s)} = D ( X , \hat{X}^{(k)} ) - D ( X , \hat{X}^{(k;s)} ) \ .
\label{eqn:RD}
\end{eqnarray}
Ideally we would like to take the next measurement at the pixel that maximizes the RD. However, because we do not know $X$, i.e. the ground-truth, the pixel that maximizes the expected reduction in distortion (ERD) is measured in the SLADS framework instead. The ERD is defined as, 
\begin{equation}
\bar{R}^{(k;s)}= \mathbb{E} \left[ R^{(k;s)} \vert Y^{(k)} \right] \ .
\label{eqn:ERD}
\end{equation}
Hence, in SLADS the goal is to measure the location, 
\begin{equation}
s^{(k+1)}= \arg \max_{s \in \left\lbrace \Omega \setminus \mathcal{S} \right\rbrace} \left\lbrace \bar{R}^{(k;s)} \right\rbrace.
\label{eqn:ideal sampling strategy}
\end{equation}

In SLADS the relationship between the measurements and the ERD for any unmeasured location $s$ is assumed to be given by, 
\begin{equation}
\mathbb{E} \left[ R^{(k;s)} \vert Y^{(k)} \right] \ = V^{(k)}_s \hat{\theta}.
\end{equation}
Here, $V^{(k)}_s$ is a $1 \times t$ feature vector extracted for location $s$ and $\hat{\theta}$ is $t \times 1$ vector that is computed in training. More details of the SLADS algorithm as well as various approximations that are necessary to make the algorithm computationally tractable are detailed in \cite{bib5}.

\section{Learning Methods}

In SLADS, the parameters to compute the ERD are learned using least-squares regression, applied on training pairs of RD due to new measurements and feature vectors from previous measurements. It is important to note that in the original SLADS algorithm \cite{bib5}, and in this paper, the RD is approximately computed to ensure training is tractable. 
In this paper, we will use support vector regression (SVR) \cite{svr} and deep neural network regression (NNR) \cite{nnr} methods to learn the parameters for ERD computation and compare their performance to classical SLADS which uses least-squares regression in training.

\subsection{Linear Least-squares}
The $ERD$ in SLADS is modeled as:
\begin{equation}
\bar{R}^{(s)} = f^{\theta}_{s}(Y) = V_s \theta
\end{equation}
The parameter vector $\theta$ is estimated by solving the least-square problem:
\begin{equation}
\hat{\theta} = \operatorname*{arg\,min}_{\theta \in \mathbb{R}^P} || {\bf R} - {\bf V}\theta ||^2
\end{equation}
Since this is an unconstrained problem, we can calculate $\theta$ by solving:
\begin{equation}
\hat{\theta} = ({\bf V}^T {\bf V})^{-1}{\bf V}^T {\bf R}
\end{equation}
where $\textbf{R}$ is an $n$-dimensional column vector formed by
\begin{equation}
\textbf{R} = 
\left[ 
\begin{array}{c}
R^{(s_1)} \\
\vdots \\
R^{(s_n)} 
\end{array} 
\right] \ ,
\label{eqn:big R training DB}
\end{equation}
and $\textbf{V}$ is given by
\begin{equation}
\textbf{V} = 
\left[ 
\begin{array}{c}
V_{s_1} \\
\vdots \\
V_{s_n} 
\end{array} 
\right] \ .
\label{eqn:big V training DB}
\end{equation}
So together $(\textbf{R},\textbf{V})$ consist of $n$ training pairs, $\{ ( R_{s_i}, V_{s_i} ) \}_{i=1}^{n}$,
that are extracted from training data during an off-line training procedure \cite{bib5}.

\subsection{Support Vector Regression}
The purpose of support vector regression is to find a smooth function that has deviation less than a tolerance $\epsilon$ of the SLADS ERDs, $\bar{R}^{(s)}$, given extracted feature vectors $V_s$:

\begin{equation}
\bar{R}^{(s)} = <w, V_s> +  b
\end{equation}
Thus, we want to minimize:
\begin{equation}
\frac{1}{2} || w ||^2 + C \sum _{i=1}^{N} (\zeta_i + \zeta_i^*)
\end{equation}
subject to:
\begin{equation}
\begin{split}
R_{s_i} - <w, V_{s_i}> - b & \leq \epsilon + \zeta_i \\
<w, V_{s_i}> + b - R_{s_i} & \leq \epsilon + \zeta_i^* \\
\zeta_i, \zeta_i^* \geq 0
\end{split}
\end{equation}
where, $\zeta_i$ and $\zeta_i^*$ are slack variables. Here, we have $C$=1, $\epsilon$ = 0.1 and Gaussian RBF kernel is used to transform the feature vector into higher dimensional feature space.

\subsection{Deep Neural Networks}
In the neural network architecture, the weights of hidden layers can be thought of as feature transformations and nodes as expanded feature vectors. The inputs to the NNR are the initial SLADS feature vectors and the outputs are the RDs. The NNR model can be defined as:
\begin{equation}
\bar{R}^{(s)} = g^{\omega} (V_s).
\end{equation}
We use function $g(\cdot)$ to denote the NNR architecture with weights between layers as $\omega$. The input for the NNR is the feature vector $V_s$ and the output is the expected reduction in distortion $\bar{R}^{(s)}$.   
The goal in NNR training is to minimize the Loss function defined as:
\begin{equation}
Loss = \frac{1}{2} \sum_{i}^{n} ( R_{s_i} - g^{\omega}(V_{s_i}) )^2
\end{equation}
Here, we have $50$ neurons in each of the 5 hidden layers in the NNR architecture and identity function as the activation function. Adam solver is used for optimization with constant learning rate of $0.001$.


\section{Experimental Data and Method}
\begin{figure}[h]
\centering
        \begin{subfigure}[htbp]{0.15\textwidth}
                \includegraphics[width=\linewidth]{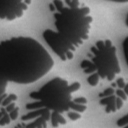}
                \caption{Training image 1.}
                \label{fig:2d}
        \end{subfigure}%
        \hspace{0.2cm}
        \begin{subfigure}[htbp]{0.15\textwidth}
                \includegraphics[width=\linewidth]{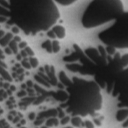}
                \caption{Training image 2.}
                \label{fig:10d}
        \end{subfigure} 
        
        \vspace{2em}
        
        \begin{subfigure}[htbp]{0.15\textwidth}
                \includegraphics[width=\linewidth]{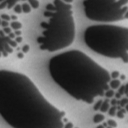}
                \caption{Test case (i).}
                \label{fig:2d}
        \end{subfigure}%
        \hspace{0.2cm}
        \begin{subfigure}[htbp]{0.15\textwidth}
                \includegraphics[width=\linewidth]{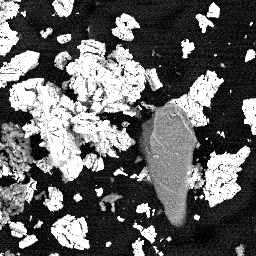}
                \caption{Test case (ii).}
                \label{fig:10d}
        \end{subfigure}%

        \vspace{1.5em}
        \caption{Images for test case (i) and (ii) \cite{bib5}\cite{yz}. (a) and (b) are used for SLADS training for each of the three learning methods. (c) is similar to the training images and used in test case (i). (d) is dissimilar to training images and used in test case (ii).}
        \label{fig:dataset}
\end{figure}

We applied the three learning methods on two scenarios: (i) when training and testing images are similar, and (ii): when training and testing images are dissimilar. 
Figure \ref{fig:dataset} shows the training and testing images used for the study here. The first set of experiments were conducted by training with images (a) and (b) and testing on image (c), which is similar to (a) and (b). The second set of experiments were conducted by training on the same images as before, but testing on image (d), which is of a different material and has 256 $\times$ 256 pixels as opposed to 128 $\times$ 128 pixels as in images (a), (b) and (c).

The purpose of these two sets of experiments was to learn if nonlinear machine learning methods achieve better results when compared to the least-squares method proposed in the original paper \cite{bib5}. 
In particular, we performed $10$ experiments on the each image, each experiment starting with a different initial set of sampling locations.
We then plotted the Peak signal-to-noise ratio (PSNR) between the ground truth images and the reconstructed images versus the sampling density to compare the methods.

\begin{figure}[h]
\centering
\vspace{1em}

        \begin{subfigure}[htbp]{0.14\textwidth}
                \includegraphics[width=\linewidth]{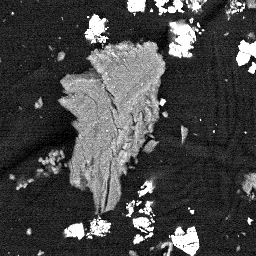}
                \caption{Pre-trained test 1.}
                \label{fig:2d}
        \end{subfigure}%
        \hspace{0.2cm}
        \begin{subfigure}[htbp]{0.14\textwidth}
                \includegraphics[width=\linewidth]{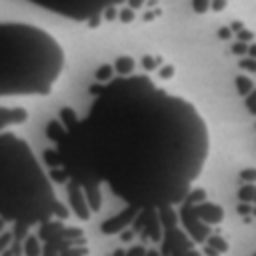}
                \caption{Pre-trained test 2.}
                \label{fig:10d}
        \end{subfigure} 
                \hspace{0.2cm}
        \begin{subfigure}[htbp]{0.14\textwidth}
                \includegraphics[width=\linewidth]{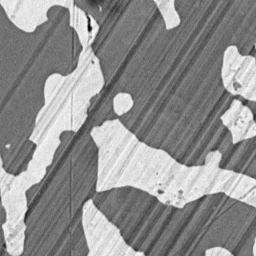}
                \caption{Pre-trained test 3.}
                \label{fig:10d}
        \end{subfigure} 
        
        
        \vspace{1.5em}
        \caption{Test images for pre-trained SLADS-Net experiment \cite{bib5}\cite{yz}.}
        \label{fig:dataset2}
\end{figure}

Then, in the final experiment, we used the generic “cameraman” image to extract feature vectors and RDs for SLADS-Net training so as to build a pre-trained SLADS-Net package.
The reason we chose the cameraman image was because it has a relatively large intensity range and a variety of textures. 
The goal of pre-trained SLADS-Net is to provide a dynamic sampling package which doesn't require training, i.e. does not require images that look similar to the sample being imaged, since such images might be hard to obtain in certain cases. 
The test images we used to evaluate the pre-trained SLADS-Net are shown in Figure \ref{fig:dataset2}. These three SEM images are from a micropowder sample, a material microstructure and a Pb-Sn alloy sample.

In SLADS training, we used the Plug \& Play algorithm \cite{plugnplay} for image reconstruction to compute the reduction-in-distortion. We used weighted mean interpolation method for image reconstruction to compute the local descriptors in order to reduce the run-time speed of SLADS.

\section{Results}
\vspace{1em}
In this section, we first discuss the results of test cases (i) and (ii) described in the previous section.
Then, we discuss the results from the pre-trained SLADS-net experiment, in which we used the cameraman image as training data. 

\subsection{Similar images in training and testing}

\begin{figure}[h]
\centering
  \includegraphics[width=0.8\columnwidth]{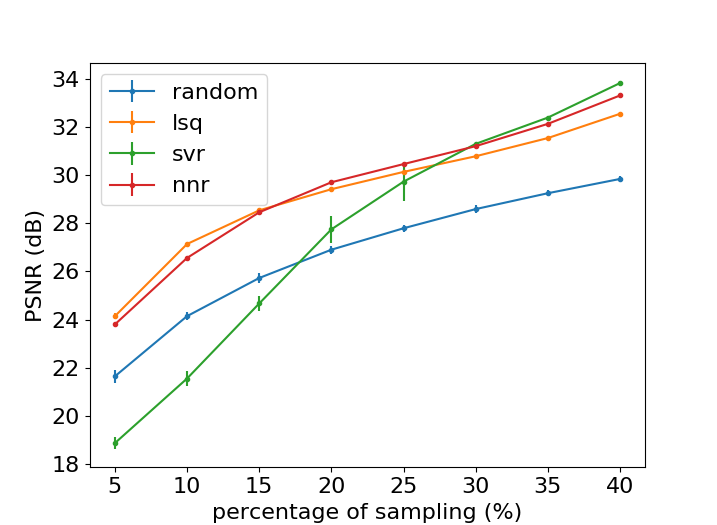}
  \vspace{1em}
  \caption{PSNR of the four methods for test case (i).}
  \label{fig:psnr1}
\end{figure}

\begin{figure*}[]
\centering
  \includegraphics[width=1.5\columnwidth]{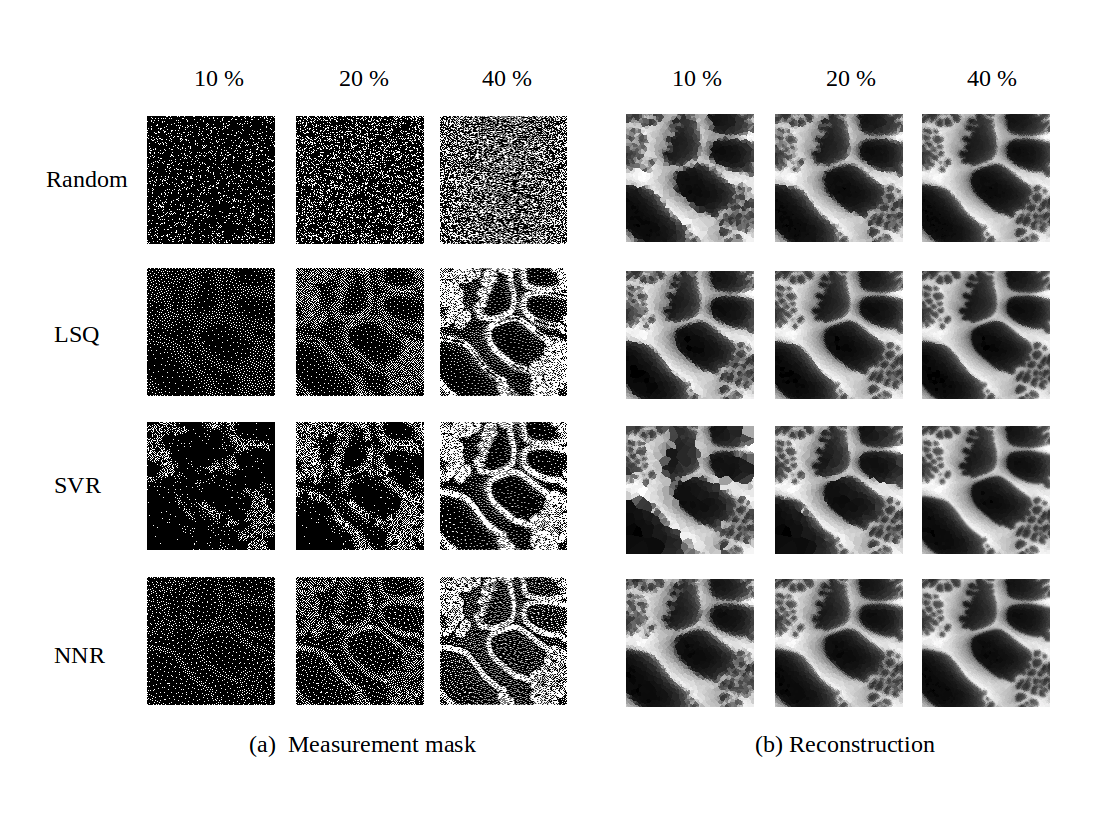}
  \caption{SLADS-Net Measurement masks and reconstructed images for test case (i). Row 1 shows random sampling results and rows 2--4 show results for the different learning methods at different sampling percentages.}
  \label{fig:mask1}
\end{figure*}

In test case (i), we observe that all learning methods achieved a high PSNR $\sim$ 33 dB when the 40\% of the test image was sampled, as shown in Table \ref{tab:psnr1}. SVR had the worst performance before 20\% sampling but had the highest PSNR of 33.83 dB at 40\% sampling, as shown in Figure \ref{fig:psnr1}. NNR had limited improvement when compared to the least-squares method in this scenario. From this result, we conclude that in the test case when training and test samples are the same, nonlinear learning methods have similar performance to the least-squares method. Measurements and reconstruction results of test case (i) are shown in Figure \ref{fig:mask1}. The last column of Table \ref{tab:psnr1} shows the average run time in test case (i): 128 $\times$ 128 image of the same samples. NNR and least-squares had much faster run time than SVR in SLADS.

\begin{table}[!h]
\vspace{2em}
\caption{PSNR and Run-Time for test case (i)}
\vspace{-1em}
\label{tab:psnr1}
\begin{center}       
\begin{tabular}{|p{0.3\columnwidth}|p{0.3\columnwidth}|p{0.2\columnwidth}|} 
\hline
Method & PSNR at 40\% & Time (s) \\ \hline \hline
random     & 29.84 $\pm$ 0.14 & N/A      \\ \hline   
lin-lsq    & 32.55 $\pm$ 0.06 & 17.59 \\ \hline
kernel-svr & 33.83 $\pm$ 0.07 & 121.68 \\ \hline
nnr    & 33.31 $\pm$ 0.05 & 16.20 \\ \hline
\end{tabular}
\end{center}
\end{table}

\subsection{Dissimilar images in training and testing}
\begin{figure}[h]
\centering
  \includegraphics[width=0.8\columnwidth]{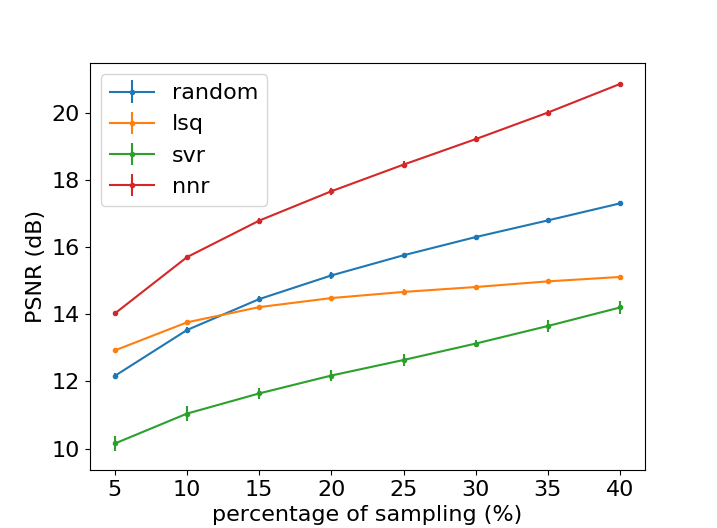}
  \vspace{1em}
  \caption{PSNR of the four methods for test case (ii).}
  \label{fig:psnr2}
\end{figure}

\begin{figure*}[!htb]
\centering
  \includegraphics[width=1.5\columnwidth]{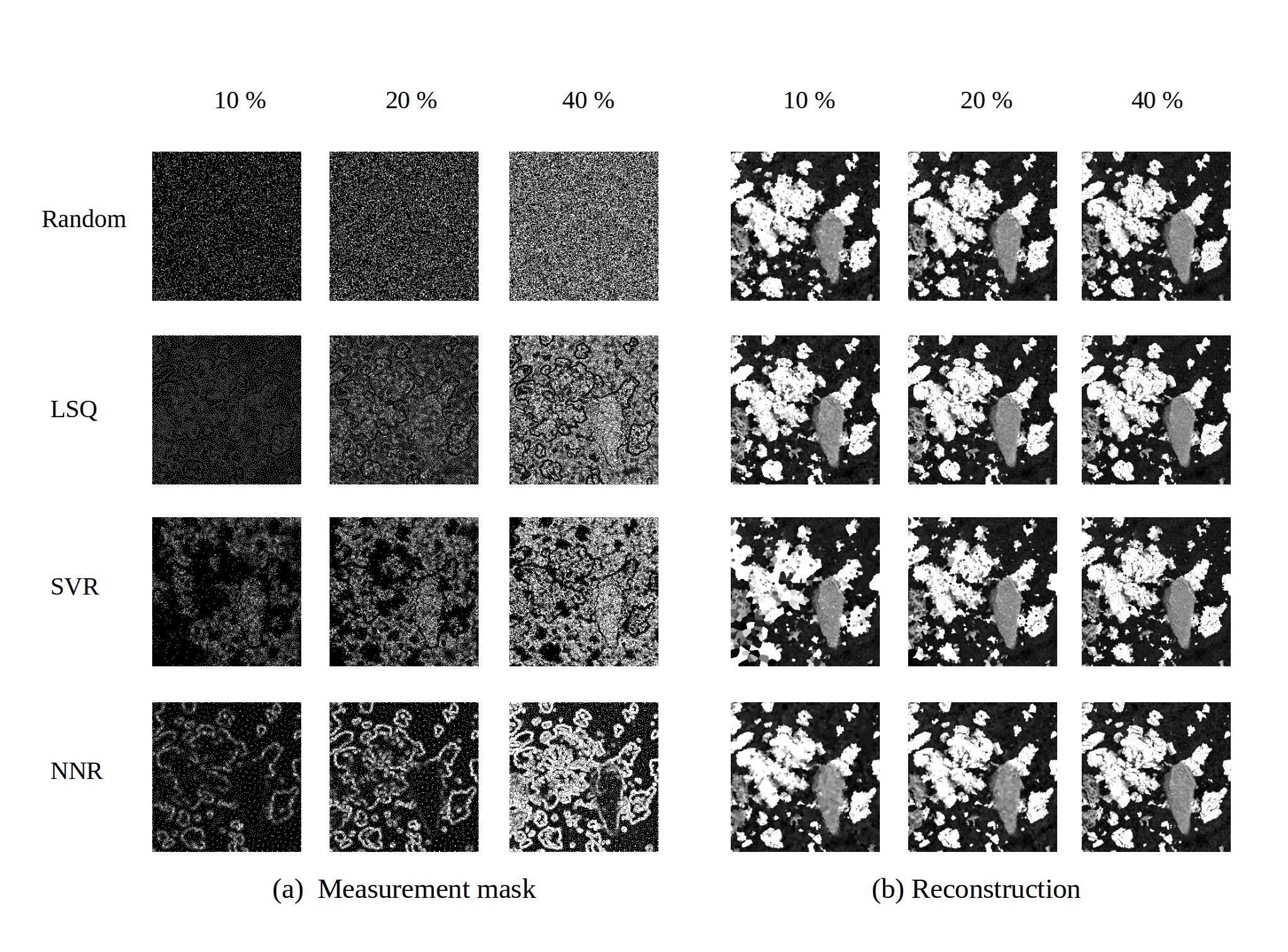}
  \caption{SLADS-Net Measurement masks and reconstructed images for test case (ii). Row 1 shows random sampling results and rows 2--4 show results for the different learning methods at different sampling percentages.}
  \label{fig:mask2}
\end{figure*}

In test case (ii), we observe that SVR and least-squares performed worse than even traditional static random sampling, as shown in Figure \ref{fig:psnr2}. We believe, this phenomenon arose because the training and test images were dissimilar. NNR on the other hand, performed well even though the parameters were learned from dissimilar images. Reconstruction using the measurements from SLADS-Net had PSNR of 20.88 dB at 40\% sampling, as shown in Table \ref{tab:psnr2}. Original SLADS and random sampling yielded PSNRs of 15.12 dB and 17.31 dB respectively at 40\% sampling. From this result, we observe that SLADS-Net dramatically outperforms original SLADS when training and test images are dissimilar. Measurements and reconstruction results of test case (ii) are shown in Figure \ref{fig:mask2}. The last column in Table \ref{tab:psnr2} shows the averaged run time in test case (ii): 256 $\times$ 256 image of different samples.
Again we observed that the SVR method was significantly slower compared to the other two methods in this test case.

\begin{table}[!h]
\vspace{2em}
\caption{PSNR and Run-Time for test case (ii)}
\vspace{-1em}
\label{tab:psnr2}
\begin{center}       
\begin{tabular}{|p{0.3\columnwidth}|p{0.3\columnwidth}|p{0.2\columnwidth}|} 
\hline
Method & PSNR at 40\% & Time (s) \\ \hline \hline
random     & 17.31 $\pm$ 0.08 & N/A      \\ \hline   
lin-lsq    & 15.12 $\pm$ 0.07 & 161.79 \\ \hline
kernel-svr & 14.21 $\pm$ 0.18 & 603.62 \\ \hline
nnr    & 20.88 $\pm$ 0.06 & 155.15 \\ \hline
\end{tabular}
\end{center}
\end{table}

\subsection{Pre-trained SLADS-Net}

\begin{figure}[!h]
\centering
        \begin{subfigure}[htbp]{0.5\textwidth}
                \includegraphics[width=\linewidth]{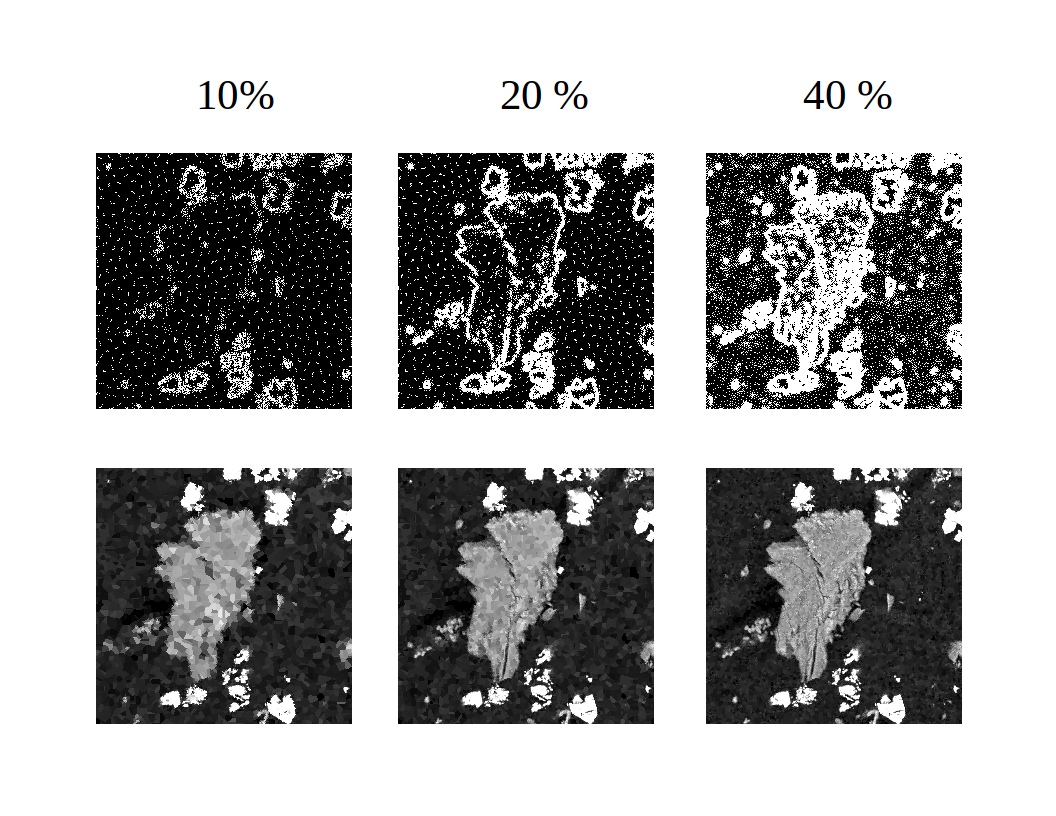}
                \vspace{-3em}
                \caption{Pre-trained SLADS-Net result for test image shown in Figure \ref{fig:dataset2}(a).}
                \vspace{1em}                
                \label{fig:2d}
        \end{subfigure} 
        
        \begin{subfigure}[htbp]{0.5\textwidth}
                \includegraphics[width=\linewidth]{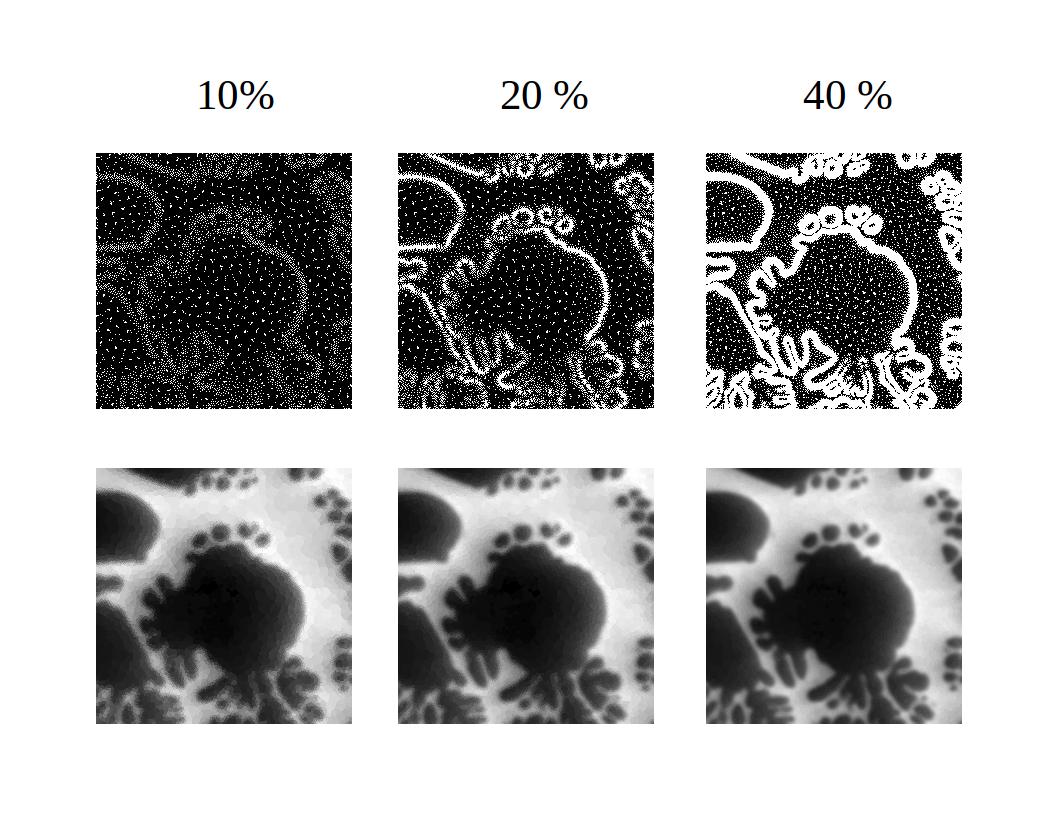}
                \vspace{-3em}
                \caption{Pre-trained SLADS-Net result for test image shown in Figure \ref{fig:dataset2}(b).}
                \vspace{1em} 
                \label{fig:10d}
        \end{subfigure} 
        
        \begin{subfigure}[htbp]{0.5\textwidth}
                \includegraphics[width=\linewidth]{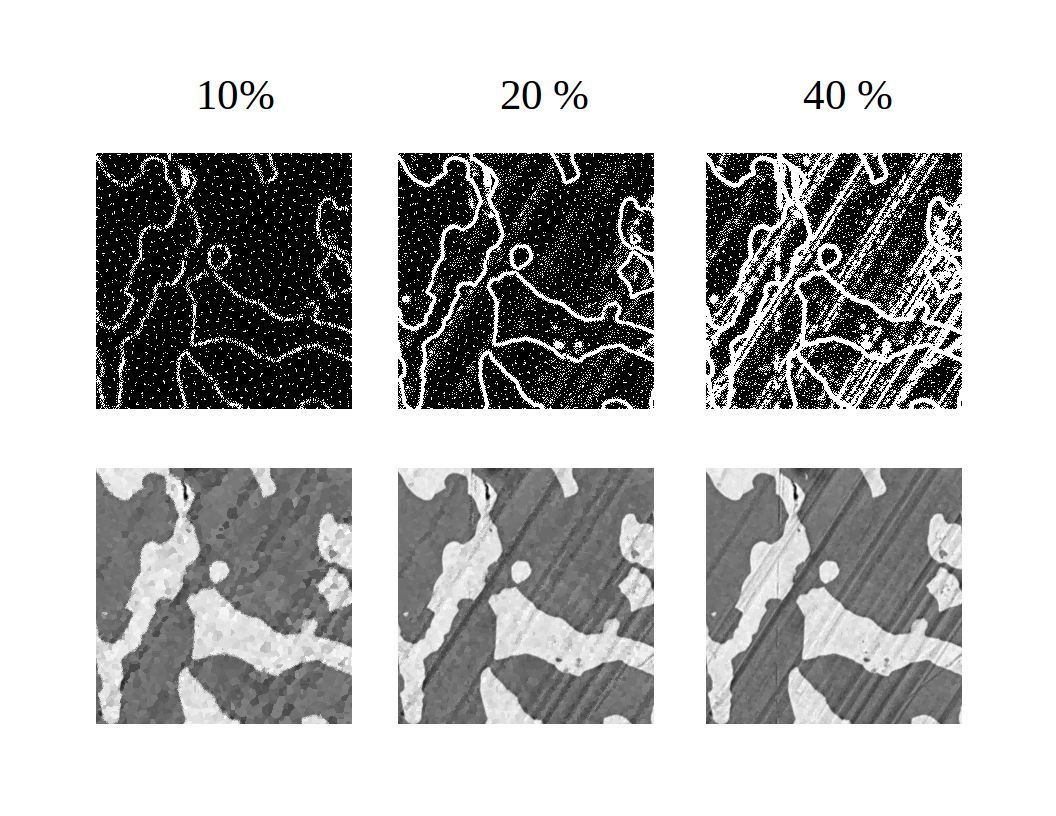}
                \vspace{-3em}
                \caption{Pre-trained SLADS-Net result for test image shown in Figure \ref{fig:dataset2}(c).}
                \vspace{1em} 
                \label{fig:10d}
        \end{subfigure} 
                
        \vspace{1.5em}
        \caption{Test results for pre-trained SLADS-Net.}
        \label{fig:maskreconcameraman}
\end{figure}

\begin{figure}[!h]
\centering
  \includegraphics[width=0.8\columnwidth]{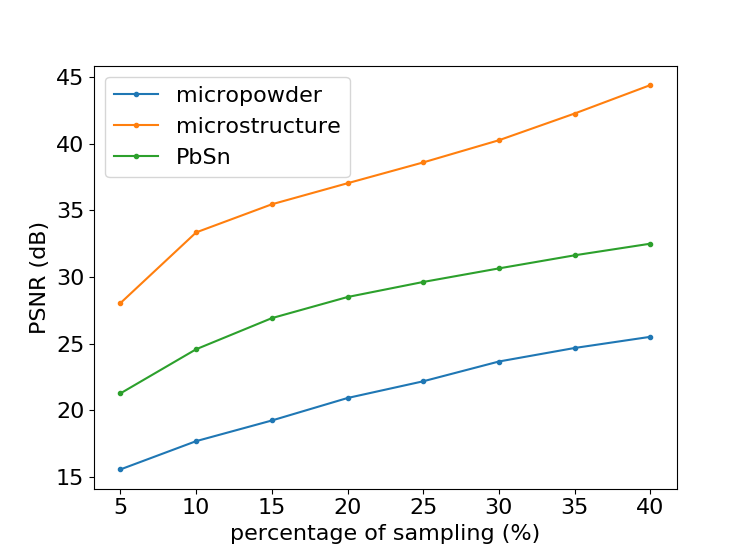}
    \vspace{1em}
  \caption{PSNR of pre-trained SLADS-Net experiment performed on test images shown in Figure \ref{fig:dataset2}.}
  \label{fig:psnr_pre}
\end{figure}

We performed the pre-trained SLADS-Net on three test images, a micropowder, a material microstructure and a Pb-Sn alloy, as shown in Figure \ref{fig:dataset2}. PSNR at 40\% sampling for the three cases were 25.51 dB, 44.40 dB and 32.50 dB when using 256 $\times$ 256 “cameraman” image in pre-trained SLADS-Net. Figure \ref{fig:maskreconcameraman} (a) to (c) shows the measurement masks and reconstructed images from 10 \% to 40 \% for the three tests. From Figure \ref{fig:psnr_pre}, we observe that test case 2 has the highest PSNR and test case 1 has the lowest PSNR. We believe, it is because test image 2 has less detailed information when compared to test image 1 which has rich features.

\section{Discussion}
Using dynamic sampling strategies to determine sampling locations can reduce the exposure time to $\sim$ 20 \%, which minimizes the sample irradiation damage for biological and beam-sensitive materials. In this paper, our goal was to compare the SLADS performance with different learning methods.
The original SLADS framework utilizes least-squares as the learning method and as a result training images need to be similar to test images for best performance.
In this paper, we proposed an improved dynamic sampling method, SLADS-Net, and a pre-trained package. SLADS-Net uses neural networks for its training phase and shows good performance even when the test sample is different from the training sample i.e. with dissimilar training and testing images.
Hence, SLADS-Net is a good solution when images similar to the test image cannot be easily found. 
As a result, pre-trained SLADS-Net using generic images enables users to apply dynamic sampling directly for imaging experiment without a training process.

\section{Acknowledgments} 
This material is based upon work supported by Laboratory Directed Research and Development (LDRD) funding from Argonne National Laboratory, provided by the Director, Office of Science, of the U.S. Department of Energy under Contract No. DE-AC02-06CH11357. Images from Figure \ref{fig:dataset} (a)-(c) and Figure \ref{fig:dataset2} (b) are provided by Ali Khosravani \& Prof. Surya Kalidindi, Georgia Institute of Technology.



\begin{biography}
Yan Zhang received his B.E. in Automation from Beijing University of Technology, China in 2010 and his Ph.D. in Electrical Engineering from Northeastern University, Boston, MA in 2016. Since 2017, he has been working as a Postdoctoral Appointee at Argonne National Laboratory, Lemont, IL.
\end{biography}

\clearpage

\end{document}